\documentclass[a4paper,11pt,twoside]{article}
\usepackage{epsfig,graphics,float}
\usepackage{subfig}
\usepackage{geometry} 
\usepackage{amsmath}
\usepackage{amssymb}
\usepackage{color}
\usepackage{graphicx}

\newcommand{\TeV}{\,\mathrm{TeV}}
\newcommand{\GeV}{\,\mathrm{GeV}}
\newcommand{\MeV}{\,\mathrm{MeV}}

\newcommand{\ord}[1]{\mathcal{O}\left( #1 \right)}
\newcommand{\vev}[1]{\left\langle #1\right\rangle}

\newcommand{\Fig}[1]{Fig.~\ref{fig:#1}}

\newcommand{\eq}[1]{eq.~(\ref{eq:#1})}

\newcommand{\eqs}[1]{eqs.~(\ref{eq:#1})}
\DeclareMathOperator{\tr}{Tr}

\newlength{\myem}
\newcommand{\sep}[1]{#1}
\newcounter{mysubequation}[equation]
\renewcommand{\themysubequation}{\alph{mysubequation}}
\newcommand{\mytag}{\stepcounter{mysubequation}%
\tag{\theequation\protect\sep{\themysubequation}}}
\newcommand{\globallabel}[1]{\refstepcounter{equation}\label{#1}}

\makeatletter
\renewcommand{\section}{\@startsection{section}{1}{0em}%
        {-3.5ex \@plus -1ex \@minus -.2ex}% 
        {2.3ex \@plus.2ex}%
        {\normalfont\large\bfseries}}
\renewcommand{\subsection}{\@startsection{subsection}{2}{0em}%
        {-3.25ex\@plus -1ex \@minus -.2ex}%
        {1.5ex \@plus .2ex}%
        {\normalfont\bfseries}}
\renewcommand{\subsubsection}%
        {\@startsection{subsubsection}{3}{0em}%
        {-3.25ex\@plus -1ex \@minus -.2ex}%
        {1.5ex \@plus .2ex}%
        {\normalfont\itshape}}
\makeatother

\newcommand{\SISSA}{SISSA/ISAS and INFN, I--34013 Trieste, Italy}

\newcommand{\preprintdate}{\hfill}
\newcommand{\preprintnumber}{%
SISSA--58/2009/EP}
 
\newcommand{\titletext}{Tree Level Gauge Mediation} 
\newcommand{\authortext}{\large Marco Nardecchia, Andrea Romanino and Robert Ziegler
\medskip\\\em\normalsize 
\SISSA}
\newcommand{\abstracttext}{We propose a new scheme in which supersymmetry breaking is communicated to the MSSM sfermions by GUT gauge interactions at the tree level. The (positive) contribution of MSSM fields to $\text{Str}(\mathcal{M}^2)$ is automatically compensated by a (negative) contribution from heavy fields. Sfermion masses are flavour universal, thus solving the supersymmetric flavour problem. In the simplest SO(10) embedding, the ratio of different sfermion masses is predicted and differs from mSugra and other schemes, thus making this framework testable at the LHC. Gaugino masses are generated at the loop level but enhanced by model dependent factors.}

\title{
\normalsize
\begin{tabular}[t]{l}%\hepnumber\\
\preprintdate\end{tabular}
\hspace*{\fill}
\begin{tabular}[t]{l}\preprintnumber\end{tabular}
\vspace{3\baselineskip}\\\Large\bfseries\titletext\bigskip}
\author{\begin{minipage}[t]{0.90\textwidth}
\normalsize\centering\authortext
\end{minipage}}
\date{}

\begin{document}

\bigskip
\maketitle
\begin{abstract}\normalsize\noindent
\abstracttext
\end{abstract}\normalsize\vspace{\baselineskip}

% \clearpage

% \noindent

Low energy supersymmetry remains one of the most attractive candidates for physics beyond the electroweak scale. While the supersymmetrization of the Standard Model (SM) is more or less straightforward, the origin of supersymmetry breaking and the mechanism through which it propagates to the SM fields and their supersymmetric partners represents the main source of theoretical uncertainty. Several options have been proposed. Among them are supergravity~\cite{GravMSB}, gauge mediation~\cite{GMSB}, anomaly mediation~\cite{AMSB}, gaugino mediation~\cite{GinoMSB}, etc. Here we consider a new option in which spontaneous supersymmetry breaking is communicated to the observable sector at the tree level through GUT (SO(10)) gauge interactions. 

Tree level supersymmetry breaking is sometimes considered not to be viable because of two issues, the supertrace formula and the fact that gaugino masses, not being generated at the tree level, turn out to be suppressed compared to sfermion masses. Both issues can be easily addressed. The supertrace formula constrains the total sfermion and the corresponding fermion total squared mass to be the same~\cite{MassSumRule}\footnote{If, as in the MSSM, there are no gauge degrees of freedom with the same quantum numbers.}. This result holds at the tree level in a globally supersymmetric theory with renormalizable K\"ahler and traceless gauge generators, which is the case of the scheme we are going to consider. This clearly represents a problem if the only fermions in chiral superfields are the SM ones, as the experimental constraints rather require a significantly larger sfermion total squared mass.

This problem is evaded in effective supergravity because the soft terms arise from non-renormalizable contributions to the K\"ahler and in standard gauge mediation (which we will sometimes call ``loop'' gauge mediation) because the soft terms arise at the loop level. In both cases one ends up with a non-vanishing supertrace. In our scheme, the supertrace does vanish (in the full theory at the GUT scale), but the positive contribution from the MSSM matter fields is automatically compensated by a negative contribution from heavier chiral superfields. In order for this to work, a SM-neutral gauge U(1) in addition to the SM hypercharge is needed to avoid a stronger implication of the supertrace formula, which requires the lightest squark in either the up or down sector not to be heavier than the corresponding lightest quark~\cite{DG}. The main features of our model arise from requiring that such an extra U(1) be part of a unified group. 

The second issue has to do with gaugino masses. Under the above hypotheses, gaugino masses can only arise at the loop level. This represents a potential problem when the sfermions are generated at the tree level because it gives rise to a hierarchy between gaugino and sfermion masses. Given the experimental bounds on the gaugino masses, $M_2\gtrsim 100\GeV$, the sfermion masses typically end up being quite heavy, $\tilde m \gtrsim (4\pi)^2 M_2/g^2 \gtrsim 10\TeV$. This would push the sfermions out of the LHC reach and would introduce a significant fine-tuning in the determination of the Higgs mass, thus approaching the split-supersymmetry regime~\cite{SplitSUSY,ADGR}. Indeed, an early implementation of some of the ideas above was considered in that context~\cite{ADGR}. As we will see, such a large hierarchy between tree level sfermion and loop gaugino masses can be avoided in our scheme because of a combination of different effects. 

In this paper we will briefly present a simple example of tree level gauge mediation, leaving a more detailed and systematic investigation to further study. 

\bigskip

\begin{figure}
\begin{center}
  \includegraphics[width=0.50\textwidth]{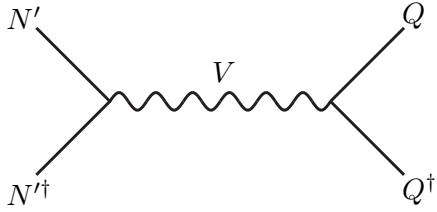}
\end{center}
  \caption{Tree level gauge mediation supergraph inducing a soft mass for the sfermion $\tilde Q$.}
\label{fig:diagram}
\end{figure}

Before presenting the model, let us motivate its gauge structure and field content. Our aim is to identify the supersymmetry breaking messengers with heavy vector superfields corresponding to broken generators, $X$, of a simple grand unified group, as illustrated in \Fig{diagram}. There, $N'$ is a SM singlet superfield whose $F$-term breaks supersymmetry, $\vev{N'} = F\,\theta^2$ (the prime is there just for consistency with the notations used below). As $N'$ has to couple to the heavy vector $V$ associated to the broken generator $X$, $N'$ must belong to a non-trivial multiplet of the unified group. $Q$ represents a generic MSSM superfield. In the effective theory below $M_\text{GUT}$, the diagram in \Fig{diagram} induces a non-renormalizable contribution $-2 g^2X_{N} X_Q (Q^\dagger Q N'^\dagger N')/M^2_V$ to the K\"ahler potential, analogous to the ones of effective supergravity, but flavour universal ($X_{N,Q}$ are the $X$-charges of $N',Q$, $M_V$ is the vector mass). A sfermion mass $\tilde m^2_Q = 2g^2 X_N X_Q (F/M_V)^2$ is then generated. In the full theory at $M_\text{GUT}$, on the other hand, everything takes place at the renormalizable level. In fact, the sfermion masses arise because $N'$ couples to the broken generator $X$. As a consequence, its $F$-term generates a non-vanishing vev for the corresponding $D$-term
\begin{equation}
\label{eq:Dterm}
\vev{D_X} = - 2 g X_{N} \bigg(\frac{F}{M_V}\bigg)^2 ,
\end{equation}
which in turn induces the soft mass
\begin{equation}
\label{eq:sfmass}
\tilde m^2_Q = - g X_Q \vev{D_X} = 2 g^2 X_{N} X_Q \bigg(\frac{F}{M_V}\bigg)^2
\end{equation}
for the sfermion $\tilde Q$. Note that there is actually no dependence on the gauge coupling (and $X$-charge normalization) because the vector squared mass $M^2_V$ is also proportional to $g^2$ (and two $X$-charges).  

Such a scheme requires specific gauge structures and field contents. First of all, the heavy vector field $V$ in \Fig{diagram} must be a SM singlet, as $N'$ is. Then, SU(5) does not provide viable candidates for the gauge messenger $V$ and the minimal option is identifying the broken generator with the SU(5) singlet generator $X$ of SO(10). As for the SM singlet $N'$ whose $F$-term breaks supersymmetry, it must belong to a non-trivial SO(10) multiplet such that $N'$ has a non-vanishing charge under $X$. Limiting ourselves to representations with dimension $d < 126$, the only possibility is that $N'$ be the singlet component of a spinorial representation, $16$ or $\overline{16}$. We also need a $16+\overline{16}$ participating to SO(10) breaking at the GUT scale. At least two $16+\overline{16}$ are then required, one getting a vev along the scalar component and the other along the $F$-term component. Gauge invariance, in fact, prevents from using a single $\vev{N'} = M + F\,\theta^2$, with both $M\neq 0$ and $F\neq 0$. This is an important difference with respect to standard gauge mediation. Finally, the standard embedding of a whole MSSM family into a 16 of SO(10) would not work, as it would lead to negative sfermion masses for some of the sfermions. That is why we distribute the matter fields in three 16 and three 10 of SO(10). 

\bigskip

Having motivated some of its features, we now illustrate a minimal model satisfying the above requirements. The gauge group is SO(10). The matter fields (negative $R$-parity) are three $16_i = (\bar 5^{16}_i, 10^{16}_i, 1^{16}_i)$ and three $10_i = (5^{10}_i, \bar 5^{10}_i)$, $i=1,2,3$, where the SU(5) decomposition is also indicated. Supersymmetry and SO(10) breaking to SU(5) are provided by $16 = (\bar 5^{16}, 10^{16}, N)$, $\overline{16} = (5^{16}, \overline{10}^{16}, \bar N)$, $16' = (\bar 5'^{16}, 10'^{16}, N')$, $\overline{16}' = (5'^{16}, \overline{10}'^{16}, \overline N')$ (positive $R$-parity), with 
\begin{equation}
\label{eq:16vev}
\vev{N'} = F\, \theta^2 \qquad \big\langle \overline{N}' \big\rangle  = 0 \qquad \vev{N} = M \qquad \big\langle \overline{N} \big\rangle = M ,
\end{equation}
$\sqrt{F} \ll M \sim M_\text{GUT}$. The $D$-term condition forces $|\langle N\rangle| = |\langle\overline{N}\rangle|$ and the phases of all the vevs can be taken positive without loss of generality. The MSSM up Higgs $h_u$ is embedded in a $10 = (5^{10},\bar 5^{10})$ of SO(10), while the down Higgs $h_d$ is a mixture of the doublets in the 10 and the 16,
\begin{equation}
\label{eq:higgsmixing}
10 = h_u + c_d h_d + \text{heavy}, \quad 16 = s_d h_d + \text{heavy} ,
\end{equation}
where $c_d = \cos\theta_d$, $s_d = \sin\theta_d$ and $0 < \theta_d < \pi/2$ parametrizes the mixing in the down Higgs sector\footnote{The most general viable Higgs embedding in this minimal model is described by the three parameters determining the up Higgs component in the 10 and the down Higgs component in the 10 and in the 16.}. We have checked that it is possible to generate such vevs, break SU(5) to the SM, achieve doublet-triplet splitting and Higgs mixing as above, and give mass to all the extra fields with an appropriate superpotential $W_\text{vev}$ involving additional SO(10) representations.

At this point we are in the condition of calculating the sfermion masses induced by integrating out the heavy vector fields: 
\begin{equation}
\label{eq:sfermions}
\tilde m^2_Q = \frac{X_Q}{2X_N}\, m^2,  \qquad m \equiv \frac{F}{M}. 
\end{equation}
In the normalization we use for $X$, $X_N = 5$. In order to determine the $X$ charge of the SM fermions we need to specify their embedding in the matter fields $16_i+10_i$. We do that by first writing the most general $R$-parity conserving superpotential, except a possible mass term for the $10_i$, as
\begin{equation}
\label{eq:W}
W = \frac{y_{ij}}{2} 16_i 16_j 10 + h_{ij} 16_i 10_j 16 + h'_{ij} 16_i 10_j 16' + W_\text{vev} + W_\text{NR} ,
\end{equation}
where $W_\text{vev} = W_\text{vev}(16,\overline{16},10,\ldots)$ does not involve the matter fields and takes care of the vevs, the doublet triplet splitting, and the Higgs mixing, and $W_\text{NR}$ contains non-renormalizable contributions to the superpotential needed in order to account for the measured ratios of down quark and charged lepton masses (we will ignore such issue here). 

We can now see that the vev of the 16 gives rise to the mass term $h_{ij}M \bar 5^{16}_i 5^{10}_j$, which makes the $\bar 5^{16}_i$ and $5^{10}_j$ heavy. Only the MSSM superfield content survives at the electroweak scale (assuming the three singlets in the $16_i$ get mass e.g.\ from non-renormalizable interactions with the $\overline{16}$). Moreover, the three MSSM families turn out to be embedded in the three $10^{16}_i$, with $X = 1$ and in the three $\bar 5^{10}_i$, with $X = 2$. We can then go back to \eq{sfermions} and obtain
\begin{gather}
\label{eq:sfermions2}
\tilde m^2_q = \tilde m^2_{u^c} = \tilde m^2_{e^c} = \tilde m^2_{10} = \frac{1}{10}\, m^2,  \quad 
\tilde m^2_{l} = \tilde m^2_{d^c} = \tilde m^2_{\bar 5} = \frac{1}{5}\, m^2 \\
\label{eq:higgses}
\quad m^2_{h_u} = -\frac{1}{5}\, m^2, \quad m^2_{h_d} = \frac{2c^2_d -3s_d^2}{10}\, m^2
\end{gather}
at the GUT scale. The result in \eq{sfermions2} is quite general, as it only depends on the choice of the gauge group and on the embedding of the three MSSM families in the $10^{16}_i + \bar 5^{10}_i$. We note a few interesting features of this result.
\begin{itemize}
\item
All the sfermion masses turn out to be positive. This is because the negative $X$ charges (which must be there as $X$ is traceless) happen to be associated to the fields that get an heavy supersymmetric mass. 
\item The sfermions masses are flavour universal, thus solving the supersymmetric flavour problem. 
\item The sfermions masses belonging to the 10 and $\bar 5$ of SU(5) are related by
\begin{equation} 
\tilde m^2_{q, u^c, e^c} = \frac{1}{2} \tilde m^2_{l, d^c}
\end{equation}
at the GUT scale, a peculiar prediction that allows to distinguish this model from mSugra, gauge mediation, and other models of supersymmetry breaking. 
\end{itemize}
Note also that the up Higgs squared mass is negative to start with, whereas $m^2_{h_d}$ is positive for $s_d < \sqrt{2/5}$. The negative value of the up Higgs squared mass means that the electroweak symmetry is broken at the tree level and the usual radiative breaking mechanism is not needed. In the presence of negative Higgs squared masses at the GUT scale, there is the potential risk that the Higgs potential develops a deep minimum along its flat direction $\tan\beta=1$, if $m^2_{h_u}+m^2_{h_d}+2|\mu|^2 < 2 |B\mu|$ at the GUT scale or below. Of course, a negative value of $m^2_{h_u}$ (and/or $m^2_{h_d}$) does not necessarily mean that  the above condition is satisfied. Moreover, in most of the parameter space, the presence of a local electroweak symmetry breaking minimum at low energy (which requires $m^2_{h_u}+m^2_{h_d}+2|\mu|^2 > 2 |B\mu|$ around the weak scale) guarantees that no deeper minima develop at higher scales. 

In passing, the SM fermion masses are given (at the renormalizable level and before running the Yukawas to low energy), by 
\begin{equation}
\label{eq:fermionmasses}
m^U_{ij} = y_{ij} v_u \qquad m^E_{ij} = \sin\theta_d h_{ij} v_d \qquad m^D_{ij} = \sin\theta_d h^T_{ij} v_d .
\end{equation}
Despite the SO(10) structure, the up quark matrix is not correlated to the down quark and charged lepton masses, which allows to accommodate the stronger mass hierarchy observed in the up quark sector. Notice that the heavy $\bar 5^{16}_i$ and $5^{10}_j$ mass matrix, $h_{ij}M$, turns out to be proportional to the charged lepton mass matrix, up to non-renormalizable corrections from $W_\text{NR}$. In the context of type-II see-saw, this can lead to a predictive model of leptogenesis~\cite{Frigerio:2008ai}. 

\bigskip

Let us now consider gaugino masses. While the tree-level prediction for the sfermion masses, \eq{sfermions2}, only depends on the choice of the unified gauge group and the MSSM embedding, gaugino masses arise at one loop, as in standard gauge mediation, and depend on the superpotential parameters. The chiral multiplets $\bar 5^{16}_i$ and $5^{10}_j$ get an heavy supersymmetric mass $h_{ij}M$ and their scalar components get a supersymmetry breaking mass $h'_{ij}F$. They play the role of three pairs of chiral messengers in standard gauge mediation and give rise to one loop gaugino masses. The contribution of each messenger arises at a different scale. In the one loop approximation for the RGE running, the total gaugino masses at lower scales can be calculated by running effective GUT-scale gaugino masses given by
\begin{equation}
\label{eq:gauginos}
M_a = \frac{\alpha}{4\pi} \tr(h'h^{-1}) \, m \equiv M_{1/2}, \quad a=1,2,3,
\end{equation}
where $\alpha$ is the unified coupling. A possible contribution from loops involving the heavy vectors vanishes (at the $F/M$ level) in this simple model. The sfermion masses also get the usual two-loop contributions. 

Let us compare gaugino and sfermion masses. Particularly interesting is the ratio $\tilde m_t / M_2$. In fact, the $W$-ino mass $M_2$ is at present bounded to be heavier than about $100\GeV$, while $\tilde m_t$ enters the radiative corrections to the Higgs mass. Therefore, the ratio $\tilde m_t / M_2$ should not be too large in order not to increase the fine-tuning and not to push the stops and the other sfermions out of the LHC reach. From 
\begin{equation}
\label{eq:ratio}
\left. \frac{M_2}{\tilde m_t} \right|_{M_\text{GUT}} = \frac{3\sqrt{10}}{(4\pi)^2}\, \lambda , \quad \lambda = \frac{g^2 \tr(h'h^{-1})}{3} 
\end{equation}
we see first of all that the loop factor separating $\tilde m_t$ and $M_2$ is partially compensated by a combination of numerical factors: $(4\pi)^2 \sim 100$ (leading to $\tilde m_t \gtrsim 10\TeV$ for $\lambda=1$) becomes $(4\pi)^2/(3\sqrt{10}) \sim 10$ (leading to $\tilde m_t \gtrsim 1\TeV$ for $\lambda=1$). Note that the factor $\sqrt{10}$ is related to the ratio of $X$ charges in \eq{sfermions}
%\footnote{More precisely the factor $\sqrt{10} = \sqrt{5\cdot 2}$ is a combination of the ratio $X(N)/X(t) = 5$ and the factor 2 corresponding to the fact that both the vevs of $N$ and $\bar N$ contribute to the vector mass suppressing sfermion masses while only the vev of $N$ is relevant to gaugino masses. In less minimal models, the latter factor 2 typically increases, thus further compensating the loop factor. That is because all the vevs breaking SO(10) contribute to the vector masses suppressing the sfermion masses (see \eq{sfmass}), while the chiral messenger masses suppressing gaugino masses only depend on fewer vevs.} 
and the factor $3$ corresponds to the number of families ($\tr(h'h^{-1}) = 3$ for $h = h'$). A largish value of the factor $\lambda$ can then further reduce the hierarchy and even make $M_2 \sim \tilde m_t$, if needed. Both $\ord{1}$ and large values of $\lambda$ are in fact not difficult to obtain depending on the overall size and flavour structure of $h$ and $h'$ (we remind that $h$ is related to the down quark Yukawa matrix and has a hierarchical structure, with two eigenvalues certainly small and the third one, related to the bottom Yukawa, also allowed to be small, depending on $\theta_d$ and $\tan\beta$). 
%Note also that the unified coupling at the GUT scale is larger than the MSSM values because of contribution to the running of the three family of chiral messengers. 

\bigskip

Reducing the hierarchy between gaugino and sfermion masses correspondingly reduces the hierarchy between the two-loop contributions to sfermion masses from standard gauge mediation and the tree level values in \eq{sfermions}. To quantify the relative importance of the two contributions, let us consider the basis in the messenger flavour space in which the matrix $h$ is diagonal and positive, the limit in which $h'$ is also diagonal in that basis, and let us call $h_i$, $h'_i$, $i=1,2,3$ their eigenvalues. Neglecting the running between the GUT scale and the mass of the relevant messengers\footnote{The relevant messengers are the ones with the largest $h'_i/h_i$. If the most relevant messenger is the third family one, the effect of the running that we are neglecting is not too large. The third family messenger mass is in fact given by $h_3 M = m_b/(v\cos\beta\sin\theta_d) M$ ($m_b$ is the bottom mass, $v=174\GeV$), not too far (in logarithmic scale) from $M\sim M_\text{GUT}$. Still, we expect the messengers to be lighter enough than the GUT scale in such a way that only the SM casimirs (and not the GUT ones) are relevant.}, the sfermion masses are given, at the high scale, by 
\begin{equation}
\label{eq:2loop}
\tilde m^2_Q = (\tilde m^2_Q)_\text{tree} + 2\, \eta \, c_Q M^2_{1/2}, \qquad
\eta = \frac{\sum (h'_i/h_i)^2}{(\sum_i h'_i/h_i)^2} \geq \frac{1}{3} ,
\end{equation}
where $(\tilde m^2_Q)_\text{tree}$ is the tree level value given in eqs.~(\ref{eq:sfermions2},\ref{eq:higgses}) and $c_Q$ is the total SM quadratic casimir of the sfermion $\tilde Q$ (or Higgs $Q$):\begin{equation}
\label{eq:casimirs}
\begin{tabular}{c|ccccccc}
$Q$ & $q_i$ & $u^c_i$ & $d^c_i$ & $l_i$ & $e^c_i$ & $h_u$ & $h_d$ \\
\hline
$c_Q$ & 21/10 & 8/5 & 7/5 & 9/10 & 3/5 & 9/10 & 9/10
\end{tabular} .
\end{equation}
If the contribution of a single messenger dominates gaugino masses, $\eta\approx 1$. In the numerical example we will consider, the relative size of the two loop contribution to sfermion masses ranges from 2\% to 10\%. 

Additional, subleading contributions to sfermion masses can arise from different sources. One-loop contributions from an induced U(1)$_X$ Fayet-Iliopoulos term~\cite{Dimopoulos:1996ig} only arise if $h'$ is non-diagonal in the basis where $h$ is diagonal and $|h'_{ij}| \neq |h'_{ji}|$. Moreover, they are suppressed (typically negligible) because U(1)$_X$ is broken above the scale of the loop messengers. Another contribution could come from gravity effects. Since in our scenario the messenger scale is expected to be around the GUT scale, the gravity mediated contribution to the spectrum, although subleading, could be relevant for flavour physics, as it could in principle be strongly flavour violating. In order to quantify this effect, let us assume that the gravity contribution to an arbitrary entry of the squared mass matrix of the sfermions in the 10 of SU(5) is given by $\tilde m^2_\text{grav} = F^2/M^2_\text{P}$, where $M_\text{P} = 2.4 \cdot 10^{18} \GeV$ is the reduced Planck mass. The conservative bound $\tilde m^2_\text{grav} < 2 \cdot 10^{-3}\, \tilde m^2_{10}$, which guarantees that all FCNC effects are under control, then translates in the following bound on the messenger scale: 
\begin{equation}
\label{eq:Mbound}
M < 3\cdot 10^{16}\GeV .
\end{equation} 
If the messenger scale is higher, we are in a hybrid framework from the flavour point of view~\cite{Hiller:2008sv}. Finally, another potentially relevant source of flavour non-universality might come from one loop contributions to sfermion masses arising from the superpotential Yukawa interactions in \eq{W}, once the (necessary) presence of mass terms for the components of the 16 and $16'$ are taken into account. Such effects are certainly under control if the matrix $h'$, as $h$, has a hierarchical structure and is approximately aligned to $h$. 

\bigskip

Let us now consider the $A$-terms. The latter are generated at one loop by the Yukawa interactions in \eq{W}, with no contribution from gauge interactions. Assuming for simplicity that the matrices $h'$ and $y$ are diagonal in the same basis in which $h$ is, we have
\globallabel{eq:Aterms}
\begin{align}
A_{l_i,d^c_i} &= 
-\frac{1}{4\pi^2} \frac{h'_i}{h_i}
\left(h^2_i + h'^2_i\right) m \mytag \\
A_{q_i,u^c_i,e^c_i} &= 
-\frac{1}{(4\pi)^2} \frac{h'_i}{h_i}
\left(
3 (h^2_i + h'^2_i) +2 y^2_i \right) m  \mytag
\end{align}
at the messenger scale. The $A$-terms above are defined in such away that they give rise to soft trilinear terms in the Lagrangian in the form $\mathcal{L} \supset -\sum_Q A_Q \tilde Q (\partial W(\tilde Q))/(\partial Q)$. By comparing with the expression for the gaugino masses, we conclude that only the $A$-terms of the third family have a chance to be sizable at the messenger scale, unless the $h'$ matrix is not hierarchical. Within the simplified diagonal flavour structure we are considering, we can compare the $A$-terms in \eqs{Aterms} with the gaugino masses in \eq{gauginos}. The gaugino masses are in this case proportional to $\sum_i h'_i/h_i$. Depending on which of the three terms dominates in the sum, the largest $A$-terms can be comparable or smaller than the gaugino masses. The (necessary) presence of mass terms for the components of the 16 and 16' can generate additional, model-dependent, contributions. In any case, sizable contributions to the $A$-terms will be generated as usual by the RGE evolution proportional to the gaugino masses. 

\bigskip

Next, we comment on the $\mu$ problem. Relating the $\mu$-term to supersymmetry breaking is, not surprisingly, a highly model-dependent issue, due to the various possibilities of implementing supersymmetry breaking and embedding the Higgs fields in SO(10). We point out, however, a simple possibility in which both the $F$-term, $\vev{N'} = F\,\theta^2$ and $\mu$ originate from the same parameter $m\sim \TeV$ in the superpotential: $W\supseteq m N' \overline N$. 

Once $\overline N$ is forced to get its vev $\vev{\overline N} = M \sim M_\text{GUT}$, $N'$ acquires an $F$-term $F = m M$ (so that $m$ is indeed the parameter introduced in \eq{sfermions}). In our setup, $N'$ and $\overline N$ are part of the SO(10) multiplets $16'$ and $\overline{16}$ respectively. A $\mu$ term related to the supersymmetry breaking scale $\mu \sim m$ is then therefore generated if $h_u$ has a component in $\overline{16}$ and $h_d$ has a component in $16'$. Such a situation can be achieved with an appropriate superpotential. Contrary to standard gauge mediation, there is no $\mu$-$B \mu$ problem here, as $B \mu / \mu$ is not enhanced by an inverse loop factor. $B \mu$ can be generated at the tree level, for example as in~\cite{ADGR}, or it can be generated by the RGE evolution. 

\bigskip

We now illustrate an example of low energy spectra that can be obtained in our framework. We neglect the (small, for our purposes) effect of the intermediate scale $\bar 5^{16}_i$ and $5^{10}_j$ and use the MSSM RGE equations, as implemented in {\tt Suspect\,2.41}~\cite{Djouadi:2002ze}, with boundary conditions at high energy as in eqs.~(\ref{eq:sfermions2},\ref{eq:higgses},\ref{eq:2loop}), the $A$-terms set to zero, and $\eta = 1$. We assume the messenger mass to coincide with the GUT scale, $M = M_\text{GUT}$. The overall normalization of the unified gaugino masses $M_{1/2}$ can be considered as a free parameter due to the presence of the factor $ \tr(h'h^{-1})$ in \eq{gauginos}, or equivalently of the factor $\lambda$ in \eq{ratio}. As the size of the parameters $\mu$ and $B \mu$ is model dependent, we consider them as free parameters as well and recover them as usual in terms of $M_Z$ and $\tan\beta$. Under the above assumptions, the parameters that specify the model are: $m$, $\theta_d$, $M_{1/2}$, $\tan \beta$ and the sign of $\mu$.

\begin{figure}
\centering
 \includegraphics[width=1.0\textwidth]{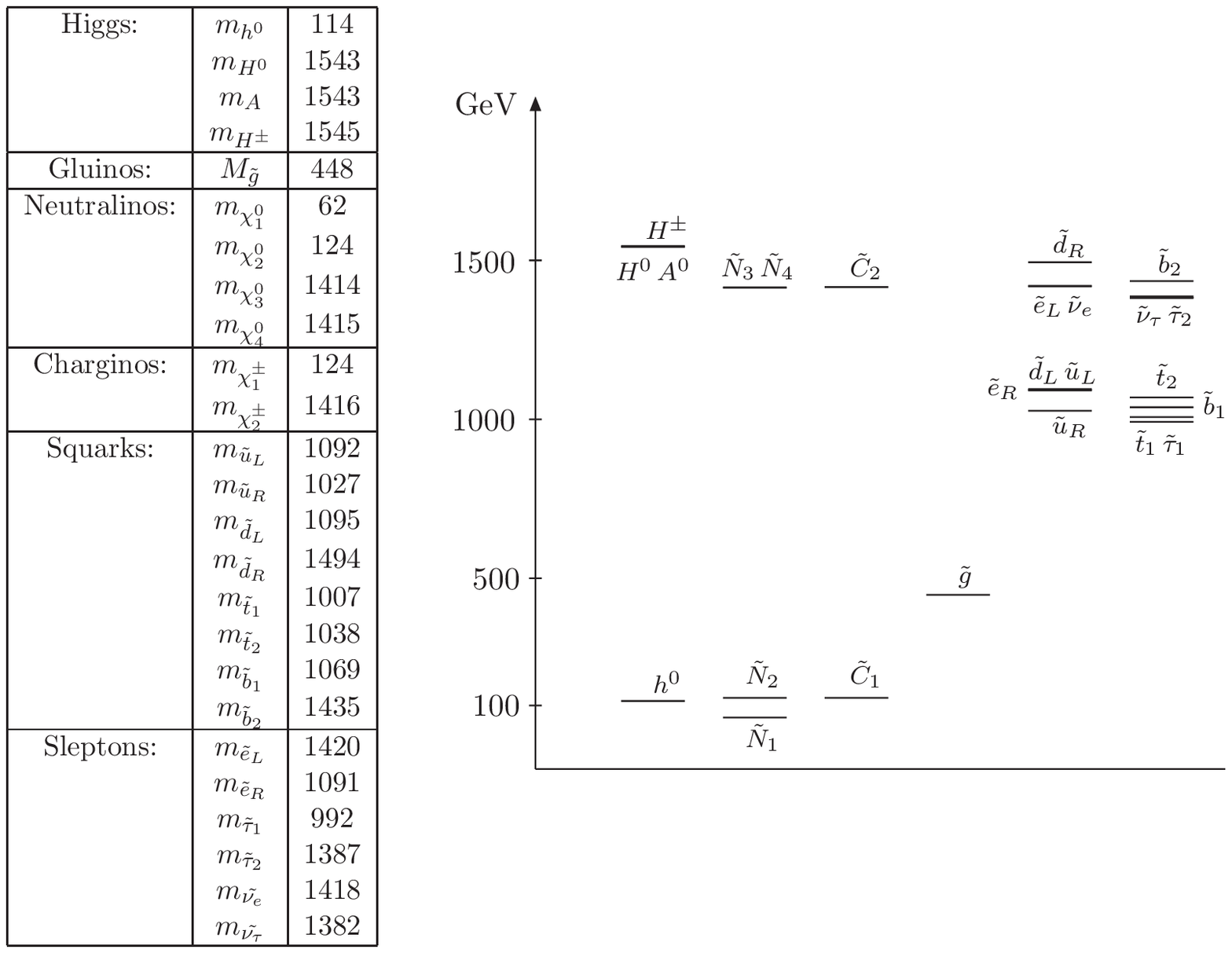}
\caption{An example of spectrum, corresponding to $m=3.2\TeV$, $M_{1/2} = 150\GeV$, $\theta_d=\pi/6$, $\tan \beta = 30$ and $\text{sign}(\mu) = +$, $A=0$, $\eta = 1$. All the masses are in $\GeV$, the first two families have an approximately equal mass.}
\label{tab:spectra}
\end{figure}

%\caption{An example of spectrum, corresponding to $m=3.2\TeV$, $M_{1/2} = 150\GeV$, $\theta_d=\pi/6$, $\tan \beta = 30$ and $\text{sign}(\mu) = +$, $A=0$, $\eta = 1$. All the masses are in $\GeV$, the first two families have an approximately equal mass.}
%\label{tab:spectra}
%\end{figure}

Table~\ref{tab:spectra} shows the low-energy spectrum corresponding to $\theta_d=\pi/6$, $\tan \beta = 30$ and $\text{sign}(\mu) = +$. The common gaugino mass is $M_{1/2} = 150\GeV$, near the minimal value allowed at present by chargino direct searches. The value of $m$ is near the minimal value allowed by the bound $m_h > 114\GeV$. This spectrum corresponds to $\lambda = 2.5$. Given the (moderate) hierarchy between $M_{1/2}$ and the sfermion masses, the sfermion RGEs are not significantly affected by the gaugino masses and the sfermion mass relations characterizing the model, \eq{sfermions2}, survive, to some extent, at low energy. The relative size of the two-loop contributions to sfermion masses in \eq{2loop} range from 2\% to 10\%. 

\bigskip

Finally, we comment about cosmology. As in loop gauge mediation, the LSP is the gravitino, if the messenger mass is consistent with \eq{Mbound}. In fact, the supersymmetry breaking parameter is given by 
\begin{equation}
\label{eq:F}
\sqrt{F} \approx 0.8 \cdot 10^{10} \GeV \bigg(\frac{\tilde m_{10}}{\TeV}\, \frac{M}{2\cdot 10^{16}\GeV}\bigg)^{1/2}
\end{equation}
and the gravitino mass by 
\begin{equation}
\label{eq:m32}
m_{3/2} = \frac{F}{\sqrt{3} M_\text{P}} \approx 15 \GeV \bigg(\frac{\tilde m_{10}}{\TeV} \,\frac{M}{2\cdot 10^{16}\GeV}\bigg) ,
\end{equation}
where $\tilde m_{10}$ is the tree-level mass of the sfermions in the 10 of SU(5) at the GUT scale.
Note that $F$ and the gravitino mass are smaller than in loop gauge mediation, for a given messenger scale $M$, because of the absence of a loop factor in eqs.~(\ref{eq:F},\ref{eq:m32}). For a stable (on the age of the universe timescale) gravitino with a mass as large as in \eq{m32}, a dilution mechanism such as inflation is necessary in order for its energy density not to exceed the dark matter one. The upper bound on the reheating temperature $T_R$ depends on the gravitino and the gaugino masses~\cite{termprod}. The thermal contribution to the gravitino energy density, for a reheating temperature around $10^9 \GeV$ is given by 
\begin{equation}
\Omega_{\tilde G}^\text{TP}h^2 \approx 6 \times 10^{-2} \left( \frac{T_{RH}}{10^9 \GeV} \right) \left( \frac{15 \GeV}{m_{3/2}} \right) \left( \frac{M_{1/2}}{150 \GeV} \right)^2 \, .
\end{equation}
For the spectrum in Table~\ref{tab:spectra}, the bound $\Omega_{\tilde G}^\text{TP}h^2 \leq \Omega_{\text{DM}} h^2 = 0.11$ translates in $T_R < 2 \cdot 10^9 \GeV $. 

We then have to take care of the decays of the NLSP into the gravitino, which might spoil big bang nucleosynthesis (BBN) unless it is fast enough. The fate of BBN depends on what the NLSP is. In the bulk of the parameter space we expect the NLSP to be the lightest neutralino or a stau. In the example in Table~\ref{tab:spectra}, the NLSP is essentially a Bino. For $m_{3/2} \sim 15\GeV$, the decay of a Bino NLSP through its coupling to the Goldstino component of the gravitino is way too slow (one would need $m_{3/2} < 100 \MeV$ in order not to spoil BBN~\cite{Feng:2004mt}). A Bino NLSP therefore requires a much faster decay channel. The latter can be provided by a tiny amount of $R$-parity violation~\cite{Buchmuller:2007ui}. Such a possibility is also consistent with thermal leptogenesis and gravitino dark matter. The other possibility is that the NLSP is a stau. In this case, all the BBN constraints can be satisfied if  the lifetime of the stau is $\tau_{\tilde{\tau}} \approx 48 \pi m_{3/2}^2 M_\text{P}^2/m^5_{\tilde{\tau}}\lesssim 6 \cdot 10^3\, s$~\cite{Pospelov:2008ta}. This is a viable possibility, which however requires large $\lambda =\ord{100}$ and sizable gaugino masses. For such large values of $\lambda$, radiative contributions to sfermion masses (from RGEs and the standard gauge mediation contribution) dominate over the tree level one, the spectrum approaches the usual loop gauge mediated one, and the peculiar relation between sfermion masses at the messenger scale gets hidden. 

\bigskip

In conclusion, we have considered what is perhaps the simplest way to communicate supersymmetry breaking: through a tree level renormalizable exchange of a gauge (GUT) messenger, as in \Fig{diagram}. We showed that this possibility is viable, despite the well known arguments associated to the supertrace formula. Besides offering new model-building avenues, this scheme solves the supersymmetric FCNC problem and, in its simplest implementation, leads to peculiar relations among sfermion masses that can be tested at the LHC. 

\section*{Acknowledgments}

We thank Matteo Bertolini, Riccardo Catena, Riccardo Rattazzi, Marco Serone, Piero Ullio for useful discussions. This work was partially supported by the RTN European Program ``UniverseNet'' (MRTN-CT-2006-035863).

%\section*{References}


\begin{thebibliography}{9}


\bibitem{GravMSB}
  H.~P.~Nilles,
  %``Dynamically Broken Supergravity And The Hierarchy Problem,''
  Phys.\ Lett.\  B {\bf 115} (1982) 193;
  %%CITATION = PHLTA,B115,193;%% 
A.H.~Chamseddine, R.~Arnowitt and P.~Nath,
  %``Locally Supersymmetric Grand Unification,''
  Phys.\ Rev.\ Lett.\  {\bf 49}, 970 (1982);
  %%CITATION = PRLTA,49,970;%%
  R.~Barbieri, S.~Ferrara and C.~A.~Savoy,
  %``Gauge Models With Spontaneously Broken Local Supersymmetry,''
  Phys.\ Lett.\ B {\bf 119}, 343 (1982);
  %%CITATION = PHLTA,B119,343;%%
  L.E.~Ib\'a\~nez, 
  %``Locally Supersymmetric SU(5) Grand Unification,''
  Phys.\ Lett.\ B {\bf 118}, 73 (1982);
  %%CITATION = PHLTA,B118,73;%%
  L.J.~Hall, J.D.~Lykken and S.~Weinberg, 
  %``Supergravity As The Messenger Of Supersymmetry Breaking,''
  Phys.\ Rev.\ D {\bf 27}, 2359 (1983);
  %%CITATION = PHRVA,D27,2359;%%
  N.~Ohta,
  %``Grand Unified Theories Based On Local Supersymmetry,''
  Prog.\ Theor.\ Phys.\  {\bf 70}, 542 (1983).
  %%CITATION = PTPKA,70,542;%% 

\bibitem{GMSB}
 M.~Dine, W.~Fischler and M.~Srednicki,
  %``Supersymmetric Technicolor,''
  Nucl.\ Phys.\  B {\bf 189}, 575 (1981);
  %%CITATION = NUPHA,B189,575;%%
  M.~Dine and W.~Fischler,
  %``A Phenomenological Model Of Particle Physics Based On
  %Supersymmetry,''
  Phys.\ Lett.\  B {\bf 110}, 227 (1982);
  %%CITATION = PHLTA,B110,227;%%
  M.~Dine and A.~E.~Nelson,
  %``Dynamical supersymmetry breaking at low-energies,''
  Phys.\ Rev.\  D {\bf 48}, 1277 (1993)
  [arXiv:hep-ph/9303230];
  %%CITATION = PHRVA,D48,1277;%%
M.~Dine, A.~E.~Nelson and Y.~Shirman,
  %``Low-Energy Dynamical Supersymmetry Breaking Simplified,''
  Phys.\ Rev.\  D {\bf 51}, 1362 (1995)
  [arXiv:hep-ph/9408384].
  %%CITATION = PHRVA,D51,1362;%%

\bibitem{AMSB}
  L.~Randall and R.~Sundrum,
  %``Out of this world supersymmetry breaking,''
  Nucl.\ Phys.\  B {\bf 557}, 79 (1999)
  [arXiv:hep-th/9810155];
  %%CITATION = NUPHA,B557,79;%%
G.~F.~Giudice, M.~A.~Luty, H.~Murayama and R.~Rattazzi,
  %``Gaugino Mass without Singlets,''
  JHEP {\bf 9812}, 027 (1998)
  [arXiv:hep-ph/9810442].
  %%CITATION = JHEPA,9812,027;%%

\bibitem{GinoMSB}
  D.~E.~Kaplan, G.~D.~Kribs and M.~Schmaltz,
  %``Supersymmetry breaking through transparent extra dimensions,''
  Phys.\ Rev.\  D {\bf 62}, 035010 (2000)
  [arXiv:hep-ph/9911293];
  %%CITATION = PHRVA,D62,035010;%%
Z.~Chacko, M.~A.~Luty, A.~E.~Nelson and E.~Ponton,
  %``Gaugino mediated supersymmetry breaking,''
  JHEP {\bf 0001}, 003 (2000)
  [arXiv:hep-ph/9911323].
  %%CITATION = JHEPA,0001,003;%%

\bibitem{MassSumRule}
%\cite{Ferrara:1979wa}
%\bibitem{Ferrara:1979wa}
  S.~Ferrara, L.~Girardello and F.~Palumbo,
  %``A General Mass Formula In Broken Supersymmetry,''
  Phys.\ Rev.\  D {\bf 20} (1979) 403.
  %%CITATION = PHRVA,D20,403;%%
  
%\cite{DG}
\bibitem{DG}
  S.~Dimopoulos and H.~Georgi,
  %``Softly Broken Supersymmetry And SU(5),''
  Nucl.\ Phys.\  B {\bf 193}, 150 (1981).
  %%CITATION = NUPHA,B193,150;%%

\bibitem{SplitSUSY} 
  N.~Arkani-Hamed and S.~Dimopoulos,
  %``Supersymmetric unification without low energy supersymmetry and  signatures
  %for fine-tuning at the LHC,''
  JHEP {\bf 0506} (2005) 073
  [arXiv:hep-th/0405159];
  %%CITATION = JHEPA,0506,073;%%
  G.~F.~Giudice and A.~Romanino,
  %``Split supersymmetry,''
  Nucl.\ Phys.\  B {\bf 699} (2004) 65
  [Erratum-ibid.\  B {\bf 706} (2005) 65]
  [arXiv:hep-ph/0406088].
  %%CITATION = NUPHA,B699,65;%%

\bibitem{ADGR}
  N.~Arkani-Hamed, S.~Dimopoulos, G.~F.~Giudice and A.~Romanino,
  %``Aspects of split supersymmetry,''
  Nucl.\ Phys.\  B {\bf 709} (2005) 3
  [arXiv:hep-ph/0409232].
  %%CITATION = NUPHA,B709,3;%%

\bibitem{Frigerio:2008ai}
  M.~Frigerio, P.~Hosteins, S.~Lavignac and A.~Romanino,
  %``A new, direct link between the baryon asymmetry and neutrino masses,''
  Nucl.\ Phys.\  B {\bf 806} (2009) 84
  [arXiv:0804.0801 [hep-ph]];
  %%CITATION = NUPHA,B806,84;%%
  L.~Calibbi, M.~Frigerio, S.~Lavignac and A.~Romanino,
  %``Flavour violation in supersymmetric SO(10) unification with a type II
  %seesaw mechanism,''
  arXiv:0910.0377 [hep-ph].
  %%CITATION = ARXIV:0910.0377;%%


%\cite{Dimopoulos:1996ig}
\bibitem{Dimopoulos:1996ig}
  S.~Dimopoulos and G.~F.~Giudice,
  %``Multi-messenger theories of gauge-mediated supersymmetry breaking,''
  Phys.\ Lett.\  B {\bf 393} (1997) 72
  [arXiv:hep-ph/9609344].
  %%CITATION = PHLTA,B393,72;%%

%\cite{Hiller:2008sv}
\bibitem{Hiller:2008sv}
  G.~Hiller, Y.~Hochberg and Y.~Nir,
  %``Flavor Changing Processes in Supersymmetric Models with Hybrid Gauge- and
  %Gravity-Mediation,''
  JHEP {\bf 0903} (2009) 115
  [arXiv:0812.0511 [hep-ph]].
  %%CITATION = JHEPA,0903,115;%%

\bibitem{Djouadi:2002ze}
  A.~Djouadi, J.~L.~Kneur and G.~Moultaka,
  ``SuSpect: A Fortran code for the supersymmetric and Higgs particle spectrum
  in the MSSM,''
  Comput.\ Phys.\ Commun.\  {\bf 176} (2007) 426
  [arXiv:hep-ph/0211331], 
  {\tt http://www.lpta.univ-montp2.fr/users/kneur/Suspect/}
  %%CITATION = CPHCB,176,426;%%

%\cite{termprod}
\bibitem{termprod}
  M.~Bolz, A.~Brandenburg and W.~Buchmuller,
  %``Thermal Production of Gravitinos,''
  Nucl.\ Phys.\  B {\bf 606} (2001) 518
  [Erratum-ibid.\  B {\bf 790} (2008) 336]
  [arXiv:hep-ph/0012052]; 
  %%CITATION = NUPHA,B606,518;%%
    V.~S.~Rychkov and A.~Strumia,
  %``Thermal production of gravitinos,''
  Phys.\ Rev.\  D {\bf 75}, 075011 (2007)
  [arXiv:hep-ph/0701104].
  %%CITATION = PHRVA,D75,075011;%% 

%\cite{Feng:2004mt}
\bibitem{Feng:2004mt}
  J.~L.~Feng, S.~Su and F.~Takayama,
  %``Supergravity with a gravitino LSP,''
  Phys.\ Rev.\  D {\bf 70} (2004) 075019
  [arXiv:hep-ph/0404231].
  %%CITATION = PHRVA,D70,075019;%%

%\cite{Buchmuller:2007ui}
\bibitem{Buchmuller:2007ui}
  W.~Buchmuller, L.~Covi, K.~Hamaguchi, A.~Ibarra and T.~Yanagida,
  %``Gravitino dark matter in R-parity breaking vacua,''
  JHEP {\bf 0703} (2007) 037
  [arXiv:hep-ph/0702184].
  %%CITATION = JHEPA,0703,037;%%

%\cite{Pospelov:2008ta}
\bibitem{Pospelov:2008ta}
  M.~Pospelov, J.~Pradler and F.~D.~Steffen,
  %``Constraints on Supersymmetric Models from Catalytic Primordial
  %Nucleosynthesis of Beryllium,''
  JCAP {\bf 0811} (2008) 020
  [arXiv:0807.4287 [hep-ph]].
  %%CITATION = JCAPA,0811,020;%
  
\end{thebibliography}
\end{document}